\documentclass[runningheads]{llncs}
\pdfoutput=1
\usepackage[T1]{fontenc}
\usepackage[utf8x]{inputenc}
\DeclareUnicodeCharacter{2227}{\ensuremath{\wedge}}
\DeclareUnicodeCharacter{2264}{\ensuremath{\leq}}
\DeclareUnicodeCharacter{2228}{\ensuremath{\vee}}
\DeclareUnicodeCharacter{2265}{\ensuremath{\geq}}
\DeclareUnicodeCharacter{22BC}{\ensuremath{\barwedge}}
\DeclareUnicodeCharacter{27F9}{\ensuremath{\Rightarrow}}
\DeclareUnicodeCharacter{27FA}{\ensuremath{\Leftrightarrow}}

\usepackage{amssymb}
\usepackage{textcomp}
\usepackage{tikz}
\usetikzlibrary {positioning, bending} 

\usepackage{booktabs}
\usepackage{color}
\usepackage[frozencache=true, cachedir=minted-cache]{minted} 

\usepackage[hidelinks]{hyperref}

\begin{document}
	\title{Satisfiability.jl:\\Satisfiability Modulo Theories in Julia\thanks{This research was supported by Ford Motor Co. under the Stanford-Ford Alliance, agreement number 235158.}
}

	%
	%
	\author{Emiko Soroka\orcidID{0009-0001-2710-469X}
		\and Mykel J. Kochenderfer\orcidID{0000-0002-7238-9663}
		\and Sanjay Lall\orcidID{0000-0002-1783-5309}
}
	\authorrunning{E. Soroka et al.}
	%
	\institute{Stanford University, Stanford CA 94305, USA\\
		\email{\{esoroka, mykel, lall\}@stanford.edu}
	}

	\maketitle              


\begin{abstract}

Satisfiability modulo theories (SMT) is a core tool in formal verification. While the SMT-LIB specification language can be used to interact with theorem proving software, a high-level interface allows for faster and easier specifications of complex SMT formulae. In this paper we present a novel open-source package for interacting with SMT-LIB compliant solvers in the Julia programming language.


\end{abstract}

\vspace{-0.5em}
\section{Introduction}
\vspace{-0.5em}
Theorem proving software is one of the core tools of formal verification, model checking, and synthesis. 
%
This paper introduces \verb|Satisfiability.jl|, a Julia package providing a high-level representation for SMT formulae including propositional logic, integer and real-valued arithmetic, and bitvectors.

Julia is a dynamically typed functional language ideal for scientific computing due to its use of type inference, multiple dispatch, and just-in-time compilation to improve performance \cite{bezanson2017julia} \cite{perkel2019juliaspeed}. 
\verb|Satisfiability.jl| provides a novel interface for SMT solving in Julia, taking advantage of language features to simplify the process of specifying SMT problems.

\subsection{Prior Work}
Many theorem provers have been developed over the years. Some notable provers include Z3 \cite{z3} and cvc5 \cite{cvc5}, both of which expose APIs in languages including C++ and Python. However, provers are low-level tools intended to be integrated into other software. Higher-level interfaces have been published for other common languages: \verb|PySMT| \cite{pysmt2015}, \verb|JavaSMT| \cite{baier2021javasmt}, and \verb|ScalaSMT| \cite{cassez2017scalasmt} are all actively maintained. In C++, the SMT-Switch library provides an interface similar to the SMT-LIB language itself \cite{mann2021smt}.

SMT solving in Julia has previously relied on wrapped C++ APIs to access specific solvers. 
However, wrapped APIs often do not match the style or best practices of a specific language. Thus an idiomatic Julia interface can make formal verification more closely integrated with Julia.

\subsection{The SMT-LIB Specification Language}
SMT-LIB is a low-level specification language designed to standardize interactions with theorem provers. At time of writing, the current SMT-LIB standard is version 2.6; we used this version of the language specification when implementing our software. To disambiguate between SMT (satisfiability modulo theories) and this specification language, we always refer to the language as SMT-LIB. Knowledge of SMT-LIB is not required to use \verb|Satisfiability.jl|. Our software provides an abstraction on top of SMT-LIB, thus we refrain from an in-depth description of the language; for a full description, readers are referred to the published standard \cite{smtlib2}. For an in-depth treatment of computational logic and the associated decision procedures, readers are referred to the books \cite{bradley2007calculus} \cite{kroening2016decision}.

\vspace{-0.5em}
\section{Package Design}
\vspace{-0.5em}
Our software provides a simple interface for solving satisfiability problems by automatically generating the underlying SMT-LIB commands and interpreting the solver responses. The basic unit of an SMT formula is the expression, which implements a tree structure capable of representing arbitrarily complex formulae.
%
%

\paragraph{Variables, expressions, and constants.}
Vector- and matrix-valued variables are arrays of single-valued expressions; thus operators can be broadcast using Julia's built in array functionality.
%
Julia's type system prevents invalid expressions from being constructed. For example, $\neg\verb|x|$ is only valid if \verb|x| is of type \verb|BoolExpr|.

The current version of our software supports propositional logic, integer and real-valued arithmetic, and bitvectors (support for other SMT-LIB theories is planned in future versions).Our \verb|BoolExpr|, \verb|IntExpr| and \verb|RealExpr| types interoperate with Julia's native \verb|Bool|, \verb|Int|, and \verb|Float64| types. Bitvector expressions interoperate with unsigned Julia integers of the appropriate size.

Expressions are simplified where possible: \verb|not(not(expr))| simplifies to \verb|expr| and nested conjunctions or disjunctions are flattened. Numeric constants are wrapped in \verb|Expr| types, simplified, and promoted as necessary: for example, adding two wrapped constants \verb|1 + 2.5| results in a single wrapped value of \verb|3.5|.

\paragraph{Uninterpreted functions.}
An uninterpreted function is a function where the input-output mapping is not known. When uninterpreted functions appear in an SMT formula, the task of the solver is to determine whether a function that satisfies the given formula exists. 
Uninterpreted functions in \verb|Satisfiability.jl| use Julia's metaprogramming capabilities to generate functions returning either SMT expressions or (if a satisfying assignment is known), the correct value when evaluating a constant.

%

\paragraph{Operators.}
Boolean and arithmetic operators are implemented as defined in SMT-LIB \cite{smtlib2}.
Design decisions were made to bring some SMT-LIB operators in line with Julia's conventions; the SMT-LIB \verb|extract| operator, which indexes into a BitVector, is represented by Julia \verb|[]| indexing, and \verb|==|, not the SMT-LIB \verb|=|, is used to construct equality constraints. Where possible, Julia symbolic operators take on their appropriate SMT-LIB meanings.

Additionally, \verb|Satisfiability.jl| provides some convenient extensions to the SMT-LIB standard syntax. Julia functions such as \verb|sum| and \verb|prod| (which call \verb|+| and \verb|*|, respectively) can be used to construct expressions. In addition to its standard meaning, the SMT-LIB \texttt{distinct} operator can accept an iterable of expressions, in which case \texttt{distinct(x1,... ,xn)} constructs a formula where each \verb|xi| must take on a unique value. Numeric types \verb|IntExpr| and \verb|RealExpr| can be mixed using Julia's type promotion functionality to call the SMT-LIB conversion functions \verb|to_int| and \verb|to_real|.  Boolean expressions can be mixed with numeric types using \verb|ite| (if-then-else): given Boolean \verb|z|, \verb|ite(z, 1 0)| converts \verb|z| to an SMT-LIB Integer and \verb|ite(z, 1.0, 0.0)| converts \verb|z| to an SMT-LIB Real. This matches the behavior of Z3.

\paragraph{Interacting with solvers.}
\vspace{-0.5em}
Internally, our package uses Julia's \verb|Process| library to interact with a solver via input and output pipes. This supports the interactive nature of SMT-LIB, which necessitates a two-way connection with the solver, and simplifies the process of making solvers available to Satisfiability.jl; the user simply ensures the solver can be invoked from their machine's command line. Users may customize the command used to invoke a solver, providing a single mechanism for customizing solver flags and interacting with any SMT-LIB solver.

\vspace{-0.5em}
\section{Usage}
\vspace{-0.5em}
\paragraph{Variables.}
New variables are declared using the \verb|@satvariable| macro, which behaves similarly to the \verb|@variable| macro in \verb|JuMP.jl| \cite{Lubin2023}.
The \verb|@satvariable| macro takes two arguments: a variable name with optional size and shape (for creating vector-valued and matrix-valued variables) and the variable type.
\vspace{-0.5em}
\begin{minted}[fontsize=\small]{julia1}
@satvariable(x, Bool) # Single BoolExpr
@satvariable(y[1:m, 1:n], Int) # m x n-vector of IntExprs
@satvariable(z[1:n], BitVector, 8) # n-vector of 8-bit BitVectorExprs
\end{minted}

\paragraph{Uninterpreted functions.}
Uninterpreted functions are declared using the
\\
\verb|@uninterpreted| macro, which generates a Julia function with the correct input and output type.
\vspace{-0.5em}
\begin{minted}[fontsize=\small]{jlcon1}
julia> @uninterpreted(f, Int, Bool)
julia> @satvariable(x, Int)
julia> sat!(¬f(-1), f(1)) # check satisfiability of a given expression
:SAT
julia> (f(-1), f(1))
(false, true)
\end{minted}

\paragraph{Operators.}
Julia's Unicode support allows many operators to be defined using their mathematical symbols. Operators represented by non-ASCII symbols may also be invoked using their ASCII text names. Operator precedence and associativity follow Julia's internal rules, described in \cite{Julia_documentation_2023}.
Users are encouraged to parenthesize expressions, as some precedence rules give unexpected results. For example:
\begin{minted}[fontsize=\small, mathescape=true]{julia1}
(x,y,z) = @satvariable(a[1:3], Bool)
x⟹y ∧ y⟹z                     # this is x $\Rightarrow$ (y $\wedge$ y $\Rightarrow$ z)
(x⟹y) ∧ (y⟹z)                 # this is (x $\Rightarrow$ y) $\wedge$ (y $\Rightarrow$ z)
and(implies(x,y), implies(y,z)) # equivalent ASCII formulation
\end{minted}

\paragraph{Working with formulae.}
The function \verb|smt(expr::AbstractExpr)| returns the SMT-LIB representation of \verb|expr| as a string. If \verb|expr| is Boolean, \verb|smt| will also assert \verb|expr|. An example is provided below.
\vspace{-0.5em}
\begin{minted}[fontsize=\small]{jlcon1}
julia> @satvariable(x, Bool)
julia> @satvariable(y, Bool)
julia> expr = or(¬x, and(¬x, y))
julia> print(smt(expr))
(declare-fun x () Bool)
(declare-fun y () Bool)
(assert (or (and (not x) y) (not x)))
\end{minted}
The function \texttt{save(e::AbstractExpr, io::IO)} writes \verb|smt(expr)| to a Julia \verb|IO| object. The function \texttt{sat!(expr::BoolExpr, s::solver)} calls the given solver on \verb|expr| and returns either \verb|:SAT|, \verb|:UNSAT|, or \verb|:ERROR|. If \verb|expr| is \verb|:SAT|, the values of all nested expressions in \verb|expr| are updated to reflect the satisfying assignment. If \verb|expr| is \verb|:UNSAT|, these values are set to \verb|nothing|.  The \verb|sat!| function can also be called on a Julia \verb|IO| object representing a string of valid SMT-LIB commands, allowing previously-written SMT-LIB files to be used with \verb|Satisfiability.jl|.

\paragraph{Interactive solving.}
SMT-LIB is an interactive interface, allowing users to modify the assumptions of an SMT problem and issue follow-up commands after a \verb|sat| or \verb|unsat| response. \verb|Satisfiability.jl| provides an \verb|InteractiveSolver| object to support these use cases. In this mode, users can manage the solver's assertion stack using \verb|push!|, \verb|pop!|, and \verb|assert!|. Calling \verb|sat!(interactive_solver)| checks the satisfiability of all current assertions, returning a tuple \texttt{(status, satisfying\_assignment)}.
%
%
%

%
Additionally, \verb|Satisfiability.jl| exposes a low-level interface, allowing advanced users to send SMT commands and receive solver responses. (Users are responsible for ensuring the correctness of these commands and interpreting the results.) Thus, \verb|Satisfiability.jl| can aid advanced SMT users by generating lengthy SMT-LIB statements and programmatically interacting with solvers.

\vspace{-0.5em}
\section{Examples}
\vspace{-0.5em}
To demonstrate the compact syntax of Satisfiability.jl,
we selected the ``Pigeonhole'' benchmark problem from the SMT-LIB benchmark library \cite{pigeon_benchmark}. Given an $(n+1)\times n$ integer matrix $P$, the problem is to find a satisfying assignment such that each element $P_{ij}$ is in $\{0,1\}$; for each row $i$, $\sum_{j=1}^{n+1} P_{ij} \geq 1$; and for each column $j$, $\sum_{i=1}^n P_{i,j} \leq 1$. (There are more rows than columns, so the problem is always unsatisfiable.) 
The benchmark is defined using the following code.
\vspace{-0.5em}
\begin{minted}[fontsize=\small]{julia1}
function pigeonhole(n::Int)
    @satvariable(P[1:n+1, 1:n], Int)
    rows = BoolExpr[sum(P[i,:]) ≥ 1 for i=1:n+1]
    cols = BoolExpr[sum(P[:,j]) ≤ 1 for j=1:n]
    status = sat!(rows, cols, P .≥ 0, P .≤ 1, solver=Z3())
    return status # should always return :UNSAT
end
\end{minted}

We also tested a graph coloring task in which \verb|Satisfiability.jl| attempts to find up to 5 colorings for each graph of size $n$, progressively adding assertions to exclude previously-found solutions. 
The following code defines the graph coloring task given $n$ (the number of nodes), a list of edges as $(i,j)$ pairs, a maximum number of colorings to find, and the number of available colors.
\vspace{-0.5em}
\begin{minted}[fontsize=\small]{julia1}
function graph_coloring(n::Int, edges, to_find::Int, colors::Int)
    @satvariable(nodes[1:n], Int)
    limits = and.(nodes .≥ 1, nodes .≤ colors)
    conns = cat([nodes[i] != nodes[j] for (i,j) in edges], dims=1)
    open(Z3()) do solver
    assert!(solver, limits, conns)
    i = 1
    while i ≤ to_find
        status, assignment = sat!(solver)
        if status == :SAT
            assign!(nodes, assignment)
            assert!(solver, not(and(nodes .== value(nodes))))
        else
            break
        end
        i += 1
    end
end
\end{minted}

\vspace{-1em}
\section{Conclusions}
\vspace{-0.5em}
We have developed a Julia package providing a simple, high-level interface to SMT-LIB compatible solvers. Our package takes advantage of Julia's functionality to construct a simple and extensible interface; we use multiple dispatch to optimize and simplify operations over constants, the type system to enforce the correctness of SMT expressions, and \verb|Base.Process| to interact with SMT-LIB compliant solvers.
\verb|Satisfiability.jl| is a registered Julia package and can be downloaded with the command \texttt{using Pkg; Pkg.add("Satisfiability.jl")}. The package documentation is at \url{https://elsoroka.github.io/Satisfiability.jl}.

\paragraph{Acknowledgments.}
The software architecture of \verb|Satisfiability.jl| was inspired by \verb|Convex.jl| and \verb|JuMP.jl| \cite{convexjl} \cite{Lubin2023}. Professor Clark Barrett provided important advice on implementing the SMT-LIB theory specifications.

%
%
%

\bibliographystyle{splncs04}    
\bibliography{bibliography}

\begin{thebibliography}{10}
\providecommand{\url}[1]{\texttt{#1}}
\providecommand{\urlprefix}{URL }
\providecommand{\doi}[1]{https://doi.org/#1}

\bibitem{Julia_documentation_2023}
 (Jul 2023),
  \url{https://docs.julialang.org/en/v1/manual/mathematical-operations/#Operator-Precedence-and-Associativity}

\bibitem{baier2021javasmt}
Baier, D., Beyer, D., Friedberger, K.: Java{SMT} 3: Interacting with {SMT}
  solvers in java. In: International Conference on Computer Aided Verification.
  pp. 195--208. Springer (2021). \doi{10.1007/978-3-030-81688-9_9}

\bibitem{cvc5}
Barbosa, H., Barrett, C.W., Brain, M., Kremer, G., Lachnitt, H., Mann, M.,
  Mohamed, A., Mohamed, M., Niemetz, A., N{\"{o}}tzli, A., Ozdemir, A.,
  Preiner, M., Reynolds, A., Sheng, Y., Tinelli, C., Zohar, Y.: cvc5: {A}
  versatile and industrial-strength {SMT} solver. In: Tools and Algorithms for
  the Construction and Analysis of Systems {TACAS} 2022. Lecture Notes in
  Computer Science, vol. 13243, pp. 415--442. Springer (2022).
  \doi{10.1007/978-3-030-99524-9\_24},
  \url{https://doi.org/10.1007/978-3-030-99524-9\_24}

\bibitem{pigeon_benchmark}
Barrett, C.: {SMT-LIB} benchmark library (Jun 2017),
  \url{https://clc-gitlab.cs.uiowa.edu:2443/SMT-LIB-benchmarks/QF_LIA/-/tree/master/pidgeons}

\bibitem{smtlib2}
Barrett, C., Fontaine, P., Tinelli, C.: {The {SMT-LIB} {S}tandard: version
  2.6}. Tech. rep., Department of Computer Science, The University of Iowa
  (2017), available at {\tt www.SMT-LIB.org}

\bibitem{bezanson2017julia}
Bezanson, J., Edelman, A., Karpinski, S., Shah, V.B.: Julia: A fresh approach
  to numerical computing. SIAM review  \textbf{59}(1),  65--98 (2017),
  \url{https://doi.org/10.1137/141000671}

\bibitem{bradley2007calculus}
Bradley, A.R., Manna, Z.: The calculus of computation: decision procedures with
  applications to verification. Springer Science \& Business Media (2007)

\bibitem{cassez2017scalasmt}
Cassez, F., Sloane, A.M.: Scala{SMT}: {S}atisfiability modulo theory in {S}cala
  (tool paper). In: Proceedings of the 8th ACM SIGPLAN International Symposium
  on Scala. pp. 51--55 (2017). \doi{3136000.3136004}

\bibitem{pysmt2015}
Gario, M., Micheli, A.: Py{SMT}: a solver-agnostic library for fast prototyping
  of {SMT}-based algorithms. In: SMT Workshop 2015 (2015)

\bibitem{kroening2016decision}
Kroening, D., Strichman, O.: Decision procedures. Springer (2016)

\bibitem{Lubin2023}
Lubin, M., Dowson, O., Garcia, J.D., Huchette, J., Legat, B., Vielma, J.P.:
  Ju{MP} 1.0: Recent improvements to a modeling language for mathematical
  optimization. Mathematical Programming Computation  (2023).
  \doi{10.1007/s12532-023-00239-3}

\bibitem{mann2021smt}
Mann, M., Wilson, A., Zohar, Y., Stuntz, L., Irfan, A., Brown, K., Donovick,
  C., Guman, A., Tinelli, C., Barrett, C.: {SMT}-switch: a solver-agnostic
  {C}++ {API} for {SMT} solving. In: International Conference on Theory and
  Applications of Satisfiability Testing. pp. 377--386. Springer (2021)

\bibitem{z3}
de~Moura, L.M., Bj{\o}rner, N.S.: Z3: An efficient {SMT} solver. In:
  International Conference on Tools and Algorithms for Construction and
  Analysis of Systems (2008). \doi{10.1007/978-3-540-78800-3_24}

\bibitem{perkel2019juliaspeed}
Perkel, J.M., et~al.: Julia: come for the syntax, stay for the speed. Nature
  \textbf{572}(7767),  141--142 (2019). \doi{10.1038/d41586-019-02310-3}

\bibitem{convexjl}
Udell, M., Mohan, K., Zeng, D., Hong, J., Diamond, S., Boyd, S.: Convex
  optimization in {J}ulia. SC14 Workshop on High Performance Technical
  Computing in Dynamic Languages  (2014). \doi{10.48550/arXiv.1410.4821}

\end{thebibliography}

\end{document}